\documentstyle[12pt]{article}
\makeatletter
\def\fnum@figure{{\hspace*{-1.5cm}Fig.\ \thefigure}}
\makeatother

\def\etal{{\em et al.}}
\def\half{{\textstyle\frac{1}{2}}}
\def\cstyle{\setlength{\baselineskip}{13pt}\advance\leftskip by 4cm
\advance\rightskip by 2.5cm}

\begin{document}

\vspace*{3.5cm}
\centerline{\large\bf PARTICLE-PARTICLE CORRELATIONS}
\centerline{\large\bf AND THE SPACE-TIME STRUCTURE}
\centerline{\large\bf OF HEAVY ION COLLISIONS}

\vspace*{1cm}
\centerline{Wolfgang BAUER}

\vspace*{0.3cm}
\centerline{National Superconducting Cyclotron Laboratory}
\centerline{and Department of Physics, Michigan State University}
\centerline{East Lansing, MI 48824-1321, USA}

\vspace*{0.3cm}
\centerline{Electronic mail address: \verb+BAUER@MSUNSCL+}

\vspace*{1cm}
\hspace*{2.5cm} ABSTRACT

The present status of the use of two-particle intensity interferometry
as a diagnostic tool to study the space-time dynamics of intermediate energy
heavy ion collisions is examined.  Calculations for the two-proton and
two-pion correlation functions are presented and compared to experiment.
The calculations are based on the nuclear Boltzmann-Uehling-Uhlenbeck
transport theory.

\vspace*{1cm}
\hspace*{2.5cm} KEYWORDS

Heavy ion transport theory; two-particle interferometry; two-pion correlation
function; two-proton correlation function; pion shadowing; nuclear equation
of state; in-medium nucleon-nucleon cross sections.

\vspace*{2cm}
\hspace*{2.5cm} INTRODUCTION

Probably the premier goal in performing heavy ion collisions at intermediate
and high beam energies is to obtain
information on the nuclear matter phase diagram, on how nuclear matter behaves
under compression and heating, and on possible phase transitions in nuclear
matter.  Trivially, the time development of the process with which we hope
to compress and heat nuclei, the collision of two heavy ions, is not directly
observable to us.  To obtain this information we are exclusively dependent
on the particles emitted during the course of the heavy ion reaction.
Some of these particles are created in the reaction process (e.g.\ photons,
pions, kaons, etas, anti-proton, $\ldots$).  Other particles have already
been present in the initial stages of the reaction (protons and neutrons) and
only have their momenta and energies modified during the collision.

All of these particles are observed a long time after the reaction has taken
place.  We are only able to observe the final four-momenta of
these particles, but not the time development that led to these final states.
Furthermore, we are not able to directly measure the location from which
these particles were emitted.  Thus we are confronted with a scenario in which
we can only observe information which is integrated over space and time.
It is our task to unfold this information to reconstruct the history of
the heavy ion reaction and to possibly extract the physically relevant
parameters of the nuclear equation of state.

In this
sense heavy ion reaction physics is like forensic medicine, in which minute
tell-tale signs and pieces of evidence left behind
are collected in order to reconstruct a crime.

The last decade has seen an extensive investigation of inclusive single
particle spectra from heavy ion collisions.  The basic idea (Stock \etal, 1982)
is that compression of nuclear matter requires energy, which is then not
available for the production of secondary particles.  By first comparing
to intranuclear cascade simulations (Stock \etal, 1982) and then to nuclear
transport calculations including mean field effects (Bertsch \etal, 1984;
Aichelin and Ko, 1985) it first seemed that there was a clear signal
for the value of the compressibility of nuclear matter.  However, further
studies including momentum dependent nuclear mean field potentials
(Aichelin \etal, 1985; Gale \etal, 1987; Gale, 1987), medium effects on the
elementary scattering process (Cugnon \etal, 1987), and improved elementary
hadron-hadron cross sections (Li and Bauer, 1991, 1991a;
Danielewicz and Bertsch, 1991) have shown that the connection between
nuclear compressibility and produced particle spectra is not as striking as
previously thought.  The present status of the theory of particle production
(Cassing \etal, 1990) indicates that there is at most a factor of 2 difference
in yield of produced particles if one changes the nuclear compressibility
constant, $\kappa$,
between values of 200 and 380 MeV.  This small sensitivity to
the nuclear compressibility is due to the fact that the amount of compressional
energy stored during the heavy ion collisions is roughly the same for all
values of $\kappa$.  Instead of storing more energy in compression, nuclei
rather adjust the maximum density reached in heavy ion collisions.

The basic problem with examining single particle inclusive cross sections is
again due to the fact that we are only observing a time and space integrated
quantity.  This problem is somewhat less severe for high energy photons,
because they are produced very early in the reaction (Bauer \etal, 1986) and
have very little final state interaction.  But pion and to some extend even
kaon spectra are strongly modified by final state interactions.
In addition, at beam energies of around 1 GeV A, the production process
of these particles can in some cases be dominated by collisions of nuclear
resonances in the nuclear medium, i.e.\ by
processes for which we do not know the
elementary hadron-hadron cross sections from experimental data.
Thus inclusive
particle production cross sections carry only limited information on the
space-time development of heavy ion reactions.

One can then ask if two-particle correlations carry more information on the
space-time history of the heavy ion reaction process.  The motivation for
this is that by observing single particles in the final state one integrates
the dynamical information over their entire world line, whereas by using
two-particle correlations one has two independent world lines, the history
of which one integrates over.
Since the relative wave function of the two particles
is modified due to their interaction at the point at which the world lines
cross or come close enough (if they do),
two-particle correlations can serve as a possible source
of information on the space time development of heavy ion reactions.

In this paper I will focus on the utilization of two-particle correlations
at small relative momenta, at which intensity interferometry is possible.

Intensity interferometry was introduced by Hanbury Brown and Twiss (1954, 1956,
1956a) as a
technique for astronomical distance measurement. They recorded
the two-photon correlation function for incoming coincident photons
as a function of their relative momentum.  This correlation function
can be written as:
\begin{equation}
  R(\vec k_1,\vec k_2) = {\langle n_{12}\rangle\over \langle n_1\rangle\langle
                               n_2\rangle} - 1\ ,
\label{eqR}
\end{equation}
where $\langle n_{12}\rangle$ is the probability of detecting two coincident
photons of wavenumber $\vec k_1$ and $\vec k_2$ in detectors 1
and 2, and $\langle n_i\rangle$ is the probability of detecting a
photon of momentum $\vec k_i$ in detector $i$ $(i=1,2)$.
Equation \ref{eqR} contains
only count rates, which are proportional to the absolute squares of
the amplitudes.  As a consequence, HBT interferometry is insensitive to phase
shifts introduced by atmospheric disturbances.  It can be used with very large
base lines and delivers superior resolution.  This was first shown
(Hanbury Brown and Twiss, 1956a) by measuring the angular diameter of Sirius.

The physical basis of the HBT
effect is that two photons have a non-zero correlation function due to the
symmetrization of their wave functions, a consequence of the quantum statistics
for identical particles.

A similar technique can also be used for source size determinations
in subatomic physics.  This was first utilized by Goldhaber \etal (1960)
in studying angular distributions of pions in $p\bar{p}$
annihilation processes.  They found that the emission probability
of coincident identical pions is strongly affected by their Bose-Einstein
statistics, which causes an enhancement of the correlation function at zero
relative momentum, $q=0$.  The width of the maximum at $q=0$
depends on the
radius of the interaction volume (Goldhaber \etal, 1960)
and also on the life-time of the emitting source (Shuryak, 1973).

In recent years, two-pion intensity interferometry has been strongly pursued
at ultra-relativistic energies, where the interest is on
the interplay between source dynamics and final state
interaction (Pratt, 1984; Kohlemainen and Gyulassy, 1986)
and pion correlations from an exploding source
(Pratt, 1986).  Intensity interferometry derives its main attraction, however,
from the prospect of its possible use as a diagnostic tool for the formation
of quark-gluon plasma (Pratt, 1986; Bertsch \etal, 1988).

Intensity interferometry is not restricted to bosons, but can also be
applied to Fermions.  Koonin (1977) proposed to use two-proton intensity
interferometry to obtain `pictures' of heavy ion collisions.
The advantage of using protons as a probe lies in the fact that they are
already present in the colliding nuclei and can be liberated relatively
easily.  In contrast, to create a pair of pions one has to spend an energy
$E_{\rm min}$ = 2\,$m_\pi$ $\approx$ 280 MeV in the center of mass of the
generating system.
Therefore, protons can be used as a probe at much lower energies.
In addition, the
two-proton relative wave function contains the prominent $^2$He-`resonance',
which leads to enhanced sensitivity of the correlation function to the source
size.  Finally, protons are easy to detect with the required
resolution.

Recent progress has been centered around the theoretical computation of
two-proton correlation functions from nuclear transport theory
(Gong \etal, 1990, 1991, 1991a; Zhu \etal, 1991, Bauer \etal, 1992, 1992a,
1992b).
In this framework, it is possible to
understand the dependence of the correlation functions on the spatial
dimension of the emitting source, on the momentum distribution of
particles in the source, and on the time development of the system
emitting the particles.
Comparisons of this theory to experimental data
have established two-proton intensity
interferometry as a quantitative tool to study heavy ion reaction dynamics.

A summary of the present status of the field as well as further references
can be found in Boal \etal\ (1990) and
Bauer \etal\ (1992).  Here I primarily focus on
the use of two proton and two pion
intensity interferometry as a diagnostic tool for
reconstruction of the space-time history of
intermediate energy heavy ion reactions and with it the
determination of limits on the parameters of interactions of nucleons in
dense nuclear matter.

\vspace{1cm}
\hspace*{2.5cm} HEAVY ION TRANSPORT THEORY FOR INTERMEDIATE ENERGIES

During the last few years, several groups have developed a nuclear
transport theory for intermediate energy (20 MeV $\leq E_{\rm beam}/A \leq$
2 GeV) heavy ion reactions
(Bertsch \etal, 1984; Kruse \etal, 1985, 1985a; Gr\'egoire \etal, 1985, 1987;
Bauer \etal, 1986; Li and Bauer, 1991; Danielewicz and Bertsch, 1991).
This transport
theory describes the time evolution of the nuclear one-body Wigner
distribution $f(\vec r,\vec p,t)$ under the influence of the nuclear
mean field and individual nucleon-nucleon collisions via
the Boltzmann-Uehling-Uhlenbeck (BUU) equation
\begin{eqnarray}
\label{BUU}
     {\partial \over \partial t} f(\vec r,\vec p,t)
    &+&{\vec p \over m} \vec\nabla_r\, f(\vec r,\vec p,t)
     -\vec\nabla_rU\, \vec\nabla_p f(\vec r,\vec p,t)
   \\
 &=&{g \over 2\pi^3\,m^2} \int \,d^3q_{1'}\,d^3q_2\,d^3q_{2'}
 \nonumber \\
 && \delta\left( {1 \over 2m}
           (p^2+q_2^2-q_{1'}^2-q_{2'}^2) \right)
    \cdot \delta^3(\vec p+\vec q_2-\vec q_{1'}-\vec q_{2'})
    \cdot {d\sigma\over d\Omega}
 \nonumber \\
 && \quad \cdot \biggl\{
    f(\vec r,\vec q_{1'},t)\,f(\vec r,\vec q_{2'},t)
    \,\Bigl(1-f(\vec r,\vec p     ,t)\Bigr)
    \,\Bigl(1-f(\vec r,\vec q_{2 },t)\Bigr)
 \nonumber \\
 && \quad \ -
    f(\vec r,\vec p     ,t)\,f(\vec r,\vec q_{2 },t)
    \,\Bigl(1-f(\vec r,\vec q_{1'},t)\Bigr)
    \,\Bigl(1-f(\vec r,\vec q_{2'},t)\Bigr) \biggr\}.
 \nonumber
\end{eqnarray}
Here $U$ is the mean field potential, which is commonly parametrized as a
density functional
\begin{equation}
  U(\rho(\vec r)) = \alpha \left(\rho(\vec r)\over\rho_0\right) +
                    \beta  \left(\rho(\vec r)\over\rho_0\right)^\sigma,
\end{equation}
where the constants $\alpha$, $\beta$, and $\sigma$ are determined from
the choice of a nuclear compressibility, $\kappa$, and from the conditions
$E/A = -15.75$ MeV and $d(E/A)/d\rho = 0$ at $\rho=\rho_0$.\ \
Popular choices are (Bertsch \etal, 1984)
$\sigma = 7/6 \Rightarrow \alpha = -358.7$ MeV, $\beta = 304.6$ MeV (`soft',
$\kappa = 200$ MeV),
$\sigma = 2 \Rightarrow \alpha = -124.4$ MeV, $\beta = 70.3$ MeV (`stiff',
$\kappa = 380$ MeV), or (Bauer \etal, 1986)
$\sigma = 4/3 \Rightarrow \alpha = -218.1$ MeV, $\beta = 164.0$ MeV (`medium',
$\kappa = 235$ MeV).
$d\sigma/d\Omega$ is the energy dependent in-medium
nucleon-nucleon scattering cross section, which is taken from free space
elementary cross section data where available, and from detailed balance and
isospin symmetry arguments otherwise.

\begin{figure}
  \vspace*{21cm}
  \caption{\cstyle Time evolution of the coordinate space density
      $\rho(x,0,z)$ in the reaction plane (in units of $\rho_0$) for the
      reaction $^{93}$Nb + $^{93}$Nb at a beam energy per nucleon
      of 60 MeV and 0 impact parameter as calculated from the solution of
      equations \protect{\ref{num1}} and \protect{\ref{num2}}.}
\end{figure}

All present solution schemes for the above equation employ the test particle
method (Wong, 1982),
in which the entire phase space is divided into small cells, whose
equations of motion are first order differential equations in time,
\begin{eqnarray}
\label{num1}
  {d\over dt} \vec p_i & = & - \vec\nabla U(\vec r_i)
      + \sum_{j\neq i} {q_i\,q_j\over (\vec r_i-\vec r_j)^2}
      + {\cal C}(\vec p_i),\\
\label{num2}
  {d\over dt} \vec r_i & = & {\vec p_i \over \sqrt{m_i^2+p_i^2}},\\
       i               & = & 1, \ldots, (A_t+A_p) {\cal N},
\end{eqnarray}
where $A_t$ and $A_p$ are the target and projectile masses, respectively, and
$\cal N$ is the number of test particles per nucleon (usually taken $>$ 100
to reduce artificially generated numerical fluctuations).
The term ${\cal C}(\vec p_i)$ represents the solution of the
collision integral via an intranuclear cascade (Cugnon \etal, 1981, 1982;
Cugnon and Lemaire, 1988) for
the test particles.  The test particle collisions respect the Pauli
exclusion principle due to the presence of the factors $(1-f)$,
which are numerically implemented via a Monte Carlo rejection method.
In our particular numerical realization the values of $f(\vec r,\vec p, t)$
are stored in a six-dimensional lattice so that the computation of the
factors $(1-f(\vec r,\vec p,t))$ only
requires the call of $2^6$ lattice elements for a six-dimensional
interpolation (Bauer, 1988).

Further details of the numerics and physics of heavy ion transport theory
can be obtained from the review articles of St\"ocker and Greiner (1986),
Bertsch and Das Gupta (1988), Schuck \etal\ (1989), Cassing and Mosel (1990),
and Wang \etal\ (1991).

The solution of equations \ref{num1} and \ref{num2} provide us with a
time evolution of the single particle phase space density $f(\vec r,\vec p,t)$
and thus yield (within the model and on a semi-classical level)
complete information
on the heavy ion reaction history.  At present this is the best and probably
only valid way to visualize the nucleus-nucleus collision process.  By
comparing the predictions of the model to experimental data one can iteratively
check the approximations entering the simulations and refine them.

For illustration we display in fig.\ 1 the time evolution of the reaction
$^{93}$Nb + $^{93}$Nb at a beam energy per nucleon of 60 MeV and 0 impact
parameter.  Displayed is the nucleon density (in units of $\rho_0$
in the reaction plane ($y$=0).  For this simulation (Bauer \etal, 1992c) a
number of ${\cal N}$=1000 test particles per nucleon was used.  The beam
direction is along the z-axis.  We can see the two nuclei touching
($t$=10 fm/c) and interpenetrating and compressing each other ($t$=40 fm/c),
an expansion phase ($t$=70-100 fm/c), and finally the formation of a
ring-shaped remnant.

It should
be noted in this context, however, that the model does not contain the
formation of complex intermediate mass fragments other than those resulting
from the decay of projectile and target remnants.  A large number of groups
presently work on the inclusion of fluctuations into the dynamics in order
to overcome this difficulty (Bauer \etal, 1987; Ayik and Gr\'egoire, 1988,
1990; Randrup and Remaud, 1990; Burgio \etal, 1991, 1992; Bonasera \etal,
1992, 1992a; Randrup, 1992; Reinhard and Suraud, 1992; Ayik \etal, 1992;
Benhassine \etal, 1992).
The verdict on the success of these attempts to describe nuclear
fragmentation is still out, and it may well be that a true N-body quantum
theory is needed.  Present attempts to approximate this theory include
transport models with dynamical production of $A\leq3$ fragments
(Danielewicz and Bertsch, 1991; Danielewicz and Pan, 1992) and on so-called
`Quantum Molecular Dynamics', semiclassical N-body simulations (Aichelin and
St\"ocker, 1986; Aichelin, 1991; Ono \etal, 1992; Peilert \etal, 1992).

\vspace{1cm}
\hspace*{2.5cm} SINGLE PARTICLE PROTON SPECTRA

Since the solution of Equation \ref{BUU} represents
the time evolution of the single particle
distribution function $f(\vec r,\vec p,t)$, it is possible
(in the limit of $t\rightarrow\infty$)
to predict all single particle observables such as proton spectra
(Aichelin and Bertsch, 1986; Bauer, 1987a) in this theory.

As one example we show in fig.\ 2 the comparison of theoretical calculations
and experimental data for the single particle inclusive proton production
cross sections.  The calculations are represented by the histograms and are
the results of the calculation of 40000 different events at random impact
parameters.  The data (Fox \etal, 1986; Chitwood \etal, 1986) are represented
by the circles.  As one can see, the overall normalization as well as
the energy and angle dependence of the experimental data are well reproduced.
There is a maximum disagreement of about a factor of 2 between calculation
and measurement.  Thus we can be confident that the phase space distribution
function integrated over coordinate space, $\int d^3r\,f(\vec r,\vec p, t)$,
reproduces the experimental observables in the limit $t\rightarrow\infty$.

\vspace{1cm}
\hspace*{2.5cm} LARGE-ANGLE CORRELATIONS BETWEEN PROTONS

Taking the calculated single-particle distributions $f(\vec r,\vec p,t)$,
we cannot make predictions for two-particle correlations.
However, if these
correlations are simply consequences of the conservation laws for momentum,
energy, angular momentum, and particle number, then it is possible
to obtain limited information on the two-particle correlation function, if
one calculates in the ensemble method.  It was shown
(Bauer, 1987a) that two-proton
correlation functions measured at large angles can
be successfully reproduced by the BUU theory, provided that total momentum
conservation is correctly taken into account.
In another investigation
Ardouin \etal\ (1988) found that the variation
of the large angle correlation function with polar angle $\theta$ is largely
due to angular momentum effects.

\begin{figure}
  \vspace*{13cm}
  \caption{\cstyle Inclusive single proton energy spectra for the reactions
      $^{16}$O + $^{12}$C at E/A = 25 MeV and
      $^{12}$O + $^{12}$C at E/A = 40 MeV.  The calculations
      (Bauer, 1987a) are represented by histograms and represent the result
      of the simulation of 40000 reactions at randomly distributed impact
      parameters.  The data (Fox \etal, 1986; Chitwood \etal, 1986) are
      represented by the circles.}
\end{figure}

\vspace{1cm}
\hspace*{2.5cm} CALCULATION OF CORRELATIONS AT SMALL RELATIVE MOMENTUM

Even though it is impossible to produce two-particle correlation functions
by only taking the calculated single particle distribution functions
$f(\vec r,\vec p,t)$, it is possible to approximately calculate the
two-particle
correlation function for small relative momenta, $\vec q = \half (\vec p_1-
\vec p_2)$, between the two emitted particles with momenta $\vec p_1$ and
$\vec p_2$.  Here the
interaction between the two particles has to be taken into account explicitly.

To derive an expression for the two-particle correlation
function, $C(\vec P,\vec q)$, we assume that
the final-state interaction between the two detected particles dominates, that
final-state interactions with the emitting source and
all remaining particles can be neglected, that
the correlation functions are determined by the two-body density of states as
corrected by the interactions between the two particles, and that the single
particle phase space distribution function of emitted particles,
$g(\vec p,x)$ varies slowly as a function of
momentum $\vec p$ (i.e. $g(\vec p,x)\approx g(\vec p \pm\vec q,x)$).
Then the  theoretical expression for the two-particle correlation
function can be written as (Koonin, 1977; Pratt, 1986; Gong \etal, 1991;
Danielewicz and Schuck, 1992)
\begin{eqnarray}
    &&
    \hspace*{-1cm} C(\vec P,\vec q) = R(\vec P, \vec q) + 1 =
    {\Pi_{12}(\vec p_1, \vec p_2) \over \Pi_1(\vec p_1)\Pi_1(\vec p_2)}
      \nonumber \\ &&
    = {{\displaystyle\int} d^4x_1 d^4x_2\,
    g({\textstyle\frac{1}{2}}\vec P,x_1) g({\textstyle\frac{1}{2}}\vec P,x_2)
    \left|\phi\left(\vec q,\vec r_1
                \!-\!\vec r_2\!+\!\frac{\vec P(t_2-t_1)}{2m}\right)\right|^2
    \over
    {\displaystyle\int} d^4x_1  \,g({\textstyle\frac{1}{2}}\vec P,x_1)
    {\displaystyle\int} d^4x_2  \,g({\textstyle\frac{1}{2}}\vec P,x_2)}\ ,
\label{eq1}
\end{eqnarray}
where $\vec P = \vec p_1 + \vec p_2$ is the total momentum of the particle
pair.  $x_1$ and $x_2$ are the space-time points of the emission of
protons 1 and 2.  $\Pi_1$ is the single- and $\Pi_{12}$ is the two-particle
emission probability.

$\phi(\vec q,\vec r)$ is the relative wave function of the particle pair.
The effect that gives rise to the HBT effect is the identical particle
interference.  In the absence of any other interaction the square of
the two particle wave function is then simply given by
\begin{equation}
   |\phi(\vec q,\vec r)|^2  = 1 \pm \cos(2\vec q\,\vec r)\ ,
\label{eq1*}
\end{equation}
where the upper sign stands for bosons and the lower for fermions.

In the presence of other interactions this result is modified.  The relative
wave function for Coulomb scattering is
\begin{equation}
  \phi_c(\vec q, \vec r) = \exp(-\half\pi\eta)\, \Gamma(1+i\eta)\, \exp(iqz)\,
      _1F_1(-i\eta|1|iq(r-z))\ ,
\end{equation}
where $_1F_1$ is the confluent hypergeometric series,
\begin{equation}
  _1F_1(a|b|z) = \sum_{k=0}^\infty {\Gamma(a+k)\, \Gamma(b)\,  z^k
                              \over \Gamma(a)\, \Gamma(b+k)\, k!}\ ,
\end{equation}
and $\eta = \alpha Z_1 Z_2 m_r / q$.
The two-pion ($\pi^+\pi^+$ or $\pi^-\pi^-$) relative wave function at small
relative momentum is usually approximated as the symmetrized Coulomb scattering
wave function
\begin{equation}
  |\phi_{\pi\pi}(\vec q,\vec r)|^2 = \half |\phi_c(\vec q, \vec r) +
                                            \phi_c(\vec q,-\vec r) |^2\ .
\end{equation}

Due to symmetries,
the wave function only depends on three independent variables
which we choose to be $q$, $r$, and $\cos\theta = \vec q\cdot\vec r/qr$.
The square of the two-pion relative wave function is displayed in fig.\ 3 as
a function of $q$ and $r$ for $\cos\theta = 0.5$.  For $r\rightarrow 0$,
$|\phi_{\pi\pi}|^2$ is given by the the Gamov penetration factor
\begin{equation}
  |\phi_{\pi\pi}(\vec q,0)|^2 = 2\, {2\pi\eta\over\exp(2\pi\eta)-1}\ .
\end{equation}

For pion pairs of opposite charge ($\pi^+\pi^-$), the wave function does not
have to be symmetrized, and it is given by the hydrogen-like
Coulomb wave function with $\eta_(\pi^+\pi^-)  = -\alpha e^2 m_r / q$.

\begin{figure}
   \vspace*{12cm}
   \caption[]{\cstyle Absolute square of the two-pion ($\pi^+\pi^+$ or
              $\pi^-\pi^-$) relative wave function,
              $|\phi_{\pi\pi}(q,r,\cos\theta\!=\!0.5)|^2$
              (From Bauer (1992b)).}
\end{figure}

For the two-proton relative wave function
the strong interaction with the prominent
$^2$He-`resonance' cannot be neglected.  To obtain the relative wave function
in this case, we solve a radial Schr\"odinger equation with the Coulomb
and the modified Reid soft core potential.  The two-proton relative wave
function is shown in fig.\ 4.  One can observe the peak at $q\approx$ 20
MeV/c due to the $^2$He-`resonance'.  Due to the effect of antisymmetrization
and due to the Coulomb interaction,
$|\phi_{pp}|\rightarrow 0$ as $r\rightarrow 0$.

Equation \ref{eq1} requires only the two-particle relative wave function and
the {\it single}-particle phase space distribution function.  Under the
assumptions stated above it is thus possible to generate {\it two}-particle
correlation functions for small relative momenta from a theory which only
predicts {\it one}-particle distribution functions.

The resulting two-particle correlation function contains information on the
space-time extension of the
emitting source.  To see this, we rewrite Equation \ref{eq1} as
\begin{equation}
  C(\vec P,\vec q) = \int d^3r\,F_{\vec P}(\vec r)\,|\phi(\vec q,\vec r)|^2\ .
\label{eq2}
\end{equation}
Here $\vec r=\vec r_1-\vec r_2$ is the relative coordinate of the two
emitted particles, and the function $F_{\vec P}(\vec r)$ is defined as
\begin{equation}
  F_{\vec P}(\vec r)={{\displaystyle\int} d^3R\,
       f({\textstyle\frac{1}{2}}\vec P,\vec R\!+\!\frac{1}{2}\vec r,t_>)
       f({\textstyle\frac{1}{2}}\vec P,\vec R\!-\!\frac{1}{2}\vec r,t_>)
                    \over
       \left({\displaystyle\int}d^3r
           f({\textstyle\frac{1}{2}}\vec P,\vec r,t_>)\right)^2}\ ,
\end{equation}
where $\vec R=\frac{1}{2}(\vec r_1+\vec r_2)$
is the center-of-mass coordinate of
the two particles, and the Wigner function $f(\vec p,\vec r,t_>)$ is the
phase space distribution of particles with momentum $\vec p$ and position
$\vec r$ at some time $t_>$ after both particles have been emitted:
\begin{equation}
      f(\vec p,\vec r,t_>) = \int_{-\infty}^{t_>} dt\,
        g(\vec p,\vec r-\vec p(t_>-t)/m,t)\ .
\end{equation}

\begin{figure}
   \vspace*{12cm}
   \caption[]{\cstyle Absolute square of the two-proton relative wave function,
              $|\phi_{pp}(q,r,\cos\theta\!=\!0.5)|^2$.
              (From Bauer (1992b))}
\end{figure}

For a given momentum $\vec P$, the correlation function has three
degrees of freedom, $\vec q$, which
are a function of
$F_{\vec P}(\vec r)$. Therefore
correlation function measurements should allow the extraction of
$F_{\vec P}(\vec r)$, the normalized probability
of two protons with the same momentum $\vec P/2$ being
separated by $\vec r$.  In this sense we are in principle
able to extract information on the coordinate space extension of the
emitting source at the emission time of the protons.

One may also use correlation function measurements to test various
theoretical models capable of predicting $g(\vec p,\vec r,t)$ and thus making
specific predictions about the correlation functions.  This approach is more
realistic in its goals, because  a full six-dimensional determination of
$C(\vec P,\vec q)$ is very difficult in practice.

\vspace{1cm}
\hspace*{2.5cm} SENSITIVITY OF THE TWO-PROTON CORRELATION FUNCTION

It is instructive to examine the sensitivity of the two-particle correlation
function to different components of the two-particle interaction.  We
perform such a study for the two-proton correlation function.
To do this, we use a simple zero-lifetime Gaussian source parameterization
\begin{equation}
  g_0(\vec p,\vec r,t) = \rho_0\,\exp(-r^2/r_0^2)\,\delta(t-t_0)\ ,
\label{eq3}
\end{equation}
where $r_0$ is the radius of the source.  Figure 5
illustrates the effects
of the different contributions to the two-proton final state interaction
for sources of different radii.  The Coulomb
interaction dominates the shape of the correlation function for
very large source radii.  For $r_0<20$ fm, the correlation
function becomes increasingly
sensitive to the effects of antisymmetrization and the strong interaction.
The strong interaction has the dominant effect for source radii around
$r_0=2.5$ fm due to the prominent $^2$He resonance.

\begin{figure}
  \vspace*{12cm}
  \caption[]{\cstyle Two-proton correlation functions calculated
             with a source
             parameterization according to Equation \ref{eq3}.  The solid
             lines represent the complete calculations including the
             effects of quantum statistics and of the Coulomb and strong
             interaction.  The dashed lines represent the case of only
             Coulomb interaction, and the dotted line is for Coulomb
             interaction plus the effect of the Fermi-Dirac statistics
             for the two protons. (From Gong \etal\ (1991))}
\end{figure}

It should be pointed out at this point, however, that in a realistic
calculation the use of a simple zero life-time Gaussian
source parametrization is not sufficient.  Instead,
calculations containing the full time and momentum dependence of
the emitting source are needed.

\vspace{1cm}
\hspace*{2.5cm} CALCULATION OF TWO-PROTON CORRELATION FUNCTIONS

We perform calculations of the single particle phase space distribution
function  $f(\vec r,\vec p, t)$ by numerically solving Equation \ref{BUU}.
These single particle distributions are then inserted into Equation \ref{eq1}
to generate the two-particle correlation function at small relative momentum.

In Figure 6, we compare our calculations of the two-proton correlation
function to experimental data for the system $^{14}$N + $^{27}$Al at
a beam energy of E/A = 75 MeV.  Since the experimental data were not
triggered on impact parameter, we have to also integrate our calculations
over impact parameter with the proper weighting factors.  (Details of this
impact parameter averaging can be found in the Appendix of Gong \etal\ (1991).)
The experimental and theoretical correlation functions are shown as a
function of the relative momentum $q$ for three different gates on the
total pair momentum $|\vec P| = |\vec p_1+\vec p_2|$.
For comparison, the beam momentum per nucleon
is $p_{\rm beam}\approx 375$ MeV/c.

\begin{figure}
  \vspace*{12cm}
  \caption[]{\cstyle Two-proton correlation function for
             the reaction $^{14}$N + $^{27}$Al at E/A = 75 MeV, as
             predicted by calculations based on the BUU theory (lines) and
             as experimentally measured (plot symbols). (From Gong \etal\
             (1991))}
\end{figure}

We show two different calculations.  The solid line represents the full
BUU calculations with the full nucleon-nucleon cross sections.  The dotted
line is the result of the BUU calculation with a reduced in-medium cross
section, $\sigma = \half \sigma_{nn}$, where $\sigma_{nn}$ is the free space
elementary nucleon-nucleon cross section.  In both cases the medium
corrections due to the Pauli-principle for the final nucleon scattering
states are, of course, taken into account.

It is clear from this figure that sizeable differences between the two
calculations with different assumptions on the in-medium nucleon-nucleon
cross section exist.  In fact, variations of the cross section by only
10\% result in differences between the calculated correlation functions
which should be experimentally measurable.  From this one has to conclude
that two-particle correlation functions are very sensitive probes for
the collisional dynamics of intermediate energy heavy ion collisions.

In fig.\ 7 we display the sensitivity of the two-proton correlation
function to the in-medium nucleon-nucleon cross section, $\sigma$,
and to the value
of the nuclear matter compressibility, $\kappa$.  We show the results of our
calculations for a stiff nuclear equation of state ($\kappa=380$ MeV) and
different values of the in-medium nucleon-nucleon cross section.

{}From our theoretical results and comparisons to experimental data we
conclude that the value of the in-medium cross section (aside from correction
due to the final state `Pauli blocking') is very close to the experimentally
measured (energy dependent) free value.  This is in agreement with
information we extracted from our investigation of the
disappearance of nuclear collective flow
(Krofcheck \etal, 1989, 1992; Ogilvie \etal, 1990; Bauer 1992a).

We also varied the compressibility of nuclear matter, which enters the
calculations through the density dependence of the mean field potential $U$.
In fig.\ 7, we also
show a calculation for a soft equation of state ($\kappa=200$ MeV).
Here we find, however, only a weak sensitivity of our results
on this parameter.  This is expected, because for the relatively light
system and low beam energy considered here only moderate maximum
values of the nuclear density are achieved.

\begin{figure}
  \vspace*{12cm}
  \caption[]{\cstyle Sensitivity of the two-proton correlation function to the
      in-medium nucleon-nucleon cross section, $\sigma$, and to the
      nuclear compressibility. (After Gong \etal\ (1990)).}
\end{figure}

\vspace{1cm}
\hspace*{2.5cm} CALCULATION OF TWO-PION CORRELATION FUNCTIONS

Pluta \etal\ (1992) have observed that when one computes the ratio
\begin{equation}
      R_{\pi^+\pi^-} = {\sigma_2(\pi^+,\pi^-) \over \sigma_1(\pi^+)
                        \sigma_1(\pi^-)}
\end{equation}
as a function of the invariant mass of the pion pair for 1.6 GeV protons
incident on heavy target nuclei, there is a striking
enhancement close to $M_{\rm inv}(\pi^+,\pi^-) = 2 M_\pi$.

\begin{figure}
  \vspace*{11cm}
  \caption[]{\cstyle Comparison of theoretical calculations (histogram,
      Li and Bauer, 1991a) and experimental data (circles Odyniec \etal, 1988)
      for single pion kinetic energy spectra from the reaction
      La+La at $E/A$ = 1.35 GeV.}
\end{figure}

We have investigated the possibility that the enhancement of the
correlation function close to an invariant mass of 2$m_\pi$ could be
caused by the effect of focussing of the two correlated pions into the
same hemisphere due to the presence of the target spectator matter,
which causes some of
the produced pions to be rescattered and absorbed.  To do
this, we performed transport calculations with an extended BUU code, which had
previously been successfully used to calculate pion spectra in heavy ion
collisions (Li \etal\ (1991)), and which has been used to show that the
observed pion collective flow (Gosset \etal\ (1991)) was caused by nuclear
shadowing effects (Li \etal\ (1991b)).

In this extended BUU code we propagate protons, neutrons, Delta and
N$^\star$ resonances, and pions.  For their interaction cross
sections, we use the experimentally measured free hadron-hadron cross
sections where available. In the cases where the elementary cross sections
are not measured we employ isospin symmetry and/or detailed balance to
obtain the unknown quantities.  The process of greatest interest here
is the creation and reabsorption of pions, which is dominated by the
two-step process
\begin{equation}
  N + N \leftrightarrow N + \Delta \leftrightarrow N + N + \pi\ .
\end{equation}

Figure 8 (Li and Bauer, 1991a)
shows the single pion spectra calculated in this model.  The reaction
is La+La $\rightarrow$ $\pi^-$+X at $E/A$ is 1.35 GeV.  The data of
Odyniec \etal\ (1988) are represented by the circles, and the results
of the calculation are shown by the histogram.  Experiment and theory
were both gated for central collisions ($b\leq 2.8$ fm), and the detection
angle for the pions was $90^\circ \pm 30^\circ$ in the center of mass
of target and projectile.  More recent calculations (Li \etal, in preparation)
use improved parametrizations for the elementary cross sections and shown
an even better agreement with experimental data.

\begin{figure}
  \vspace*{12cm}
  \caption{\cstyle Two pion ($\pi^+,\pi^-$) correlation function as
      a function of the invariant mass of the pion pair for p + C, Pb
      reaction at an incident energy of 1.6 GeV.  The data are from
      Pluta \etal\ (1992) and are represented by the squares.  Our
      calculations (Klakow \etal\ (1992)) are represented by the histograms.}
\end{figure}

To generate correlation functions which are comparable to experiment,
we calculate the correlated pairs from events with
fixed reaction plane
\begin{equation}
   M_{\rm inv} = \sqrt{(E_1+E_2)^2 - (\vec p_1 + \vec p_2)^2}
\end{equation}
whereas we compute the uncorrelated background events from
\begin{equation}
   M_{\rm inv,\phi} = \sqrt{(E_1+E_2)^2 - (\vec p_1 + {\cal R}_\phi
                                           \vec p_2)^2}
\end{equation}
where $E_i$ and $\vec p_i$ are the energy and momentum of pion $i$ in
the some frame of reference (we chose the lab frame).  The operator
${\cal R}_\phi$ rotates a vector $\vec p$ by a (random) angle $\phi$
about the beam axis (the $z$-axis in our case):
\begin{equation}
   {\cal R}_\phi \vec p = \left(
      \begin{array}{ccc} \cos\phi & -\sin\phi & 0 \\
                         \sin\phi &  \cos\phi & 0 \\
                         0        &  0        & 1
      \end{array}\right) \ \left(
      \begin{array}{c} p_x \\ p_y \\ p_z \end{array}\right)
\end{equation}

The correlation function is then computed from
\begin{equation}
  R(M_{\rm inv}) = { N(M_{\rm inv}) \over
		     N(\langle M_{\rm inv,\phi} \rangle_\phi)}
\end{equation}

With this technique we approximate the experimental fact that for the
generation of the background events one is forced to average over
random relative orientations of the reaction planes for the two
uncorrelated pions in every pair, whereas the correlated pairs are all
sampled with fixed reaction plane of the pair and therefore contain the
effects of the target matter shadowing at finite impact parameter.
For a more detailed description of the numerical procedure and the
effect of pion absorption on two pion correlation functions see
Klakow \etal\ (1992).

In fig.\ 9, we compare the results of our calculations with the
experimental data for the C and Pb targets.  We find that we can
reproduce the Pb target data by the calculations, whereas we slightly
overpredict the correlation function at small $M_{\rm inv}$ for the C
target.

The main result of this investigation is thus that the effect of shadowing
due to pion absorption has a large effect on the extracted two-pion
correlation function, particularly in asymmetric projectile-target systems.
As shown this kinematic effect leads to an enhancement of the two-pion
correlation function at small $M_{\rm inv}$ and can lead to wrong
extracted source sizes if not included.

\vspace{1cm}
\hspace*{2.5cm} CONCLUSIONS

The calculation of two-particle correlation functions at small
relative momentum on the basis of
one-body transport theories is feasible by using the convolution techniques
described above.  Thus intensity interferometry is a powerful tool to
test nuclear transport theories and to investigate nuclear dynamics.

Comparisons with experiment show that the BUU transport theory is able
to reproduce detailed features of the experimentally measured
two-proton correlation functions.  These features include the pair momentum
dependence of the peak due to the $^2$He-`resonance', the effect of source
deformation, and the lifetime effect on the correlation function.

We have shown that the theoretically obtained two-proton
correlation functions are sensitive to the value of the in-medium
nucleon-nucleon cross section.  At higher beam energies and for large
systems we also expect
a sensitivity of the results on the compressibility of nuclear matter.
Thus nuclear intensity interferometry is a useful tool to investigate
the nuclear transport properties, and it should also enable us to conduct
further studies of the nuclear equation of state.

{}From a theoretical standpoint it is clearly desirable to compare
to impact parameter resolved experimental data, which will further increase
the sensitivity of two-particle correlation functions at small relative
momentum to the effects discussed
above.  These studies are currently in progress.

For the two-pion correlation functions we have shown that it is essential
to included the effects of pion rescattering and absorption properly into
the calculations.  A source characterization without taking these effects
into account will lead to wrong results for quantities like the size of
the emitting source.

For the next few years, we anticipate that the focus of HBT type of studies
in heavy ion collisions will also
shift to the energy range of $E/A \approx$ 1 GeV.
At this energy range we expect much higher compression ($\rho/\rho_0
\approx 2-4$) of the colliding nuclei than at the beam energy range
considered here ($E/A \approx 100$ MeV, and $\rho/\rho_0 \approx 1-3-1.5$).
Then two-particle correlations
should be able to give us additional information on the
nuclear compressibility and thus provide valuable insight into the nuclear
equation of state.  Work in this direction is alrady in progress
(Mader and Bauer, 1992).

\vspace{1cm}
\hspace*{2.5cm} ACKNOWLEDGMENTS

This work was supported by the National Science Foundation under Grants No.\
89-06116 and 90-17077 as well as by a Presidential Faculty Fellow award.
Helpful conversations and collaborations with
G.F. Bertsch, P. Danielewicz, C.K. Gelbke, W.G. Gong, D. Klakow,
C. Mader, S. Pratt, and P.Schuck are gratefully acknowledged.

\vspace{1cm}
\hspace*{2.5cm} REFERENCES
\vspace*{0.3cm}

{
\advance\leftskip by 0.5cm
\parindent=-0.5cm
\parskip=0pt

Aichelin, J. and Bertsch, G.F. (1985). Numerical simulation of medium
      energy heavy ion reactions.  {\em Phys.\ Rev.\ C} {\bf 31}, 1730.

Aichelin, J. and Ko, C.M. (1985). Subthreshold kaon production as a probe of
      the nuclear equation of state.  {\em Phys.\ Rev.\ Lett.} {\bf 55}, 2661.

Aichelin, J. \etal\ (1985). Importance of momentum-dependent interactions
      for the extraction of the nuclear equation of state from high-energy
      heavy-ion collisions.  {\em Phys.\ Rev.\ Lett.} {\bf 58}, 1926.

Aichelin, J. and St\"ocker, H. (1986).  Quantum molecular dynamics - A novel
      approach to N-body correlations in heavy ion collisions.
      {\em Phys.\ Lett.} {\bf B176}, 14.

Aichelin, J. (1991). ``Quantum'' molecular dynamics - a dynamical microscopic
      n-body approach to investigate fragment formation and the nuclear
      equation of state in heavy ion collisions.
      {\em Phys.\ Rep.} {\bf 202}, 233.

Ardouin, D. \etal\ (1988).  Evidence for persisting mean field effects at E/A
      = 60 MeV from particle-particle correlation measurements and
      theoretical investigations with the Landau Vlasov equation.
      {\em Z.\ Phys.\ A} {\bf 329}, 505.

Ayik, S. and Gr\'egoire, C. (1988).  Fluctuations of single-particle density
      in nuclear collisions.
      {\em Phys.\ Lett.} {\bf B212}, 269.

Ayik, S. and Gr\'egoire, C. (1990).  Transport theory of fluctuation phenomena
      in nuclear collisions.
      {\em Nucl.\ Phys.\ A} {\bf 513}, 187.

Ayik, S. \etal\ (1992).  The Boltzmann-Langevin model for nuclear collisions.
      {\em Nucl.\ Phys.\ A} {\bf 545}, 35c.

Bauer, W. \etal\ (1986). Energetic photons from intermediate energy proton
      and heavy ion induced reactions.  {\em Phys.\ Rev.\ C} {\bf 34}, 2127.

Bauer, W. \etal\ (1987). Fluctuations and clustering in heavy ion
      collisions. {\em Phys.\ Rev.\ Lett.} {\bf 58}, 863.

Bauer, W. (1987a). Light particle correlations in heavy ion reactions.
      {\em Nucl.\ Phys.\ A} {\bf 471}, 604 .

Bauer, W. (1988). Nuclear stopping at intermediate beam energies.
      {\em Phys.\ Rev.\ Lett.} {\bf 61}, 2534.

Bauer, W. (1989). Unified calculation of photon and pion spectra in
      intermediate energy heavy ion reactions.
      {\em Phys.\ Rev.\ C} {\bf 40}, 715 (1989).

Bauer, W. \etal\ (1992). Hadronic interferometry in heavy ion collisions.
      {\em Annu.\ Rev.\ Nucl.\ Part.\ Sci.} {\bf 42}, 77.

Bauer, W. (1992a). Hadronic transport properties in intermediate energy heavy
      ion collisions.  {\em Nucl.\ Phys.} {\bf A538}, 83c.

Bauer, W. (1992b). Nuclear dynamics and intensity interferometry.
      {\em Nucl.\ Phys.} {\bf A545}, 369c.

Bauer, W. \etal\ (1992c).  Bubble and ring formation in nuclear fragmentation.
      {\em Phys.\ Rev.\ Lett.} {\bf 69}, 1888.

Benhassine, B. \etal\ (1992). Phase space fluctuations and dynamics of
      fluctuations of collective variables.
      {\em Nucl.\ Phys.\ A} {\bf 545}, 81c.

Bertsch, G.F. \etal\ (1984). Boltzmann equation for heavy ion collisions.
      {\em Phys.\ Rev.\ C} {\bf 29}, 673.

Bertsch, G.F. \etal\ (1988). Pion interferometry in ultrarelativistic
      heavy-ion collisions. {\em Phys.\ Rev.\ C} {\bf 37}, 1896.

Bertsch, G.F. and Das Gupta, S. (1988).  A guide to microscopic models for
      intermediate energy heavy ion collisions.
      {\em Phys.\ Rep.} {\bf 160}, 189.

Boal, D.H. \etal\ (1990). Intensity Interferometry in subatomic physics.
      {\em Rev.\ of Mod.\ Phys.} {\bf 62}, 553.

Bonasera, A. \etal\ (1992). Fluctuations of the one-body distribution function.
      {\em Phys.\ Rev.\ C} {\bf 46}, 1431.

Bonasera, A. \etal\ (1992a). Quasistationary description of fluctuations.
      {\em Nucl.\ Phys.\ A} {\bf 545}, 71c.

Burgio, F. \etal\ (1991).  Fluctuations in nuclear dynamics.  From transport
      theory to dynamical simulation.
      {\em Nucl.\ Phys.\ A} {\bf 529}, 157.

Burgio, F. \etal\ (1991). Dynamical clusterization in the presence of
      instabilities.  {\em Phys.\ Rev.\ Lett.} {\bf 69}, 885.

Cassing, W. \etal\ (1990). Production of energetic particles in heavy-ion
      collisions.  {\em Phys.\ Rep.} {\bf 188}, 363.

Cassing, W. and Mosel, U. (1990).  Many-body theory of high-energy
      heavy-ion reactions.
      {\em Prog.\ Part.\ Nucl.\ Phys.} {\bf 25}, 235.

Chitwood, C.B. \etal\ (1986).  Light particle emission in $^{16}$O-induced
      reactions on $^{12}$C, $^{27}$Al, and $^{197}$Au at E/A = 25 MeV.
      {\em Phys.\ Rev.\ C} {\bf 34}, 858.

Cugnon, J. \etal\ (1981). Equilibration in relativistic nuclear collisions.
      A Monte Carlo calculation.
      {\em Nucl.\ Phys.\ A} {\bf 352}, 505.

Cugnon, J. \etal\ (1982). Pion production in central high energy nuclear
      collisions.
      {\em Nucl.\ Phys.\ A} {\bf 379}, 553.

Cugnon, J. \etal\ (1987). Medium effects in the nuclear Landau-Vlasov transport
      theory. {\em Phys.\ Rev.\ C} {\bf 35}, 861.

Cugnon, J. and Lemaire, M.C. (1988). Medium effects in pion production.
      {\em Nucl.\ Phys.\ A} {\bf 489}, 781.

Danielewicz, P. and Bertsch, G.F. (1991).  Production of deuterons and pions
      in a transport model of energetic heavy ion reactions.
      {\em Nucl.\ Phys.\ A} {\bf 533}, 712.

Danielewicz, P. and Schuck, P. (1992).  Formulation of particle correlation
      and cluster production in heavy-ion induced reactions.
      {\em Phys.\ Lett.} {\bf B274}, 268.

Danielewicz, P. and Pan, Q. (1992).  Blast of light fragments from central
      heavy-ion collisions.
      {\em Phys.\ Rev.\ C} {\bf 46}, Number 5.

Fox, D. \etal\ (1986).  Large angle correlations in 40 MeV/nucleon
      $^{12}$C + C.
      {\em Phys.\ Rev.\ C} {\bf 33}, 1540.

Gale, C. (1987). Pion multiplicities in dynamical models of heavy ion
      collisions. {\em Phys.\ Rev.\ C} {\bf 36}, 2152.

Gale, C. \etal\ (1987). Heavy-ion collision theory with momentum-dependent
      interactions.  {\em Phys.\ Rev.\ C} {\bf 35}, 1666.

Goldhaber, G. \etal\ (1960).  Influence of Bose-Einstein statistics on the
      antiproton-proton annihilation process.  {\em Phys.\ Rev.} {\bf 120},
300.

Gong, W.G. \etal\ (1990). Intensity interferometric test of nuclear collision
      geometries obtained from the Boltzmann-Uehling-Uhlenbeck equation.
      {\em Phys.\ Rev.\ Lett.} {\bf 65}, 2114.

Gong, W.G. \etal\ (1991). Space-time evolution of nuclear reactions probed
      by two-proton intensity interferometry. {\em Phys.\ Rev.\ C}
      {\bf 43}, 781.

Gong, W.G. \etal\ (1991a). Space time evolution of the reactions
      $^{14}$N + $^{27}$Al,
      $^{197}$Au, at E/A = 75 MeV and $^{129}$Xe + $^{27}$Al, $^{122}$Sn
      at E/A = 31 MeV probed by two-proton intensity interferometry.
      {\em Phys.\ Rev.\ C} {\bf 43}, 1804.

Gosset, J. \etal\ (1989).  Nuclear collective flow and charged pion emission
      in Ne-nucleus collisions at E/A = 800 MeV.
      {\em Phys.\ Rev.\ Lett.} {\bf 62}, 1251.

Gr\'egoire, C. \etal\ (1985). Semiclassical approaches to proton emission
      in intermediate-energy heavy-ion reactions.
      {\em Nucl.\ Phys.\ A} {\bf 436}, 365.

Gr\'egoire, C. \etal\ (1987). Semiclassical dynamics of heavy-ion reactions.
      {\em Nucl.\ Phys.\ A} {\bf 465}, 317.

Hanbury Brown, R., and Twiss, R.Q. (1954). {\em Phil.\ Mag.} {\bf 45}, 663.

Hanbury Brown, R., and Twiss, R.Q. (1956). {\em Nature} {\bf 177}, 27.

Hanbury Brown, R., and Twiss, R.Q. (1956). A test of a new type of stellar
      interferometer on sirius. {\em Nature} {\bf 178}, 1046.

Li, B.A. and Bauer, W. (1991). Two-temperature shape of pion spectra in
      relativistic heavy ion collisions. {\em Phys.\ Lett.} {\bf B254}, 335.

Li, B.A. and Bauer, W. (1991a). Pion spectra in a hadronic transport model
      for relativistic heavy ion collisions.  {\em Phys.\ Rev.\ C} {\bf 44},
      450.

Li, B.A. \etal\ (1991b). Preferential emission of pions in asymmetric
      nucleus-nucleus collisions.
      {\em Phys.\ Rev.\ C} {\bf 44}, 2095.

Klakow, D. (1992) Work in preparation.

Koonin, S.E. (1977). Proton pictures of high-energy nuclear collisions.
      {\em Phys.\ Lett.} {\bf B70}, 43 (1977).

Krofcheck, D. \etal\ (1989). Disappearance of flow in heavy ion collisions.
      {\em Phys.\ Rev.\ Lett.} {\bf 63}, 2028.

Krofcheck, D. \etal\ (1992). Disappearance of flow as a probe of the nuclear
      equation of state.
      {\em Phys.\ Rev.\ C} {\bf 46}, 1416.

Kruse, H. \etal\ (1985).  Microscopic theory of pion production and sidewards
      flow in heavy-ion collisions.
      {\em Phys.\ Rev.\ Lett.} {\bf 54}, 289.

Kruse, H. \etal\ (1985a).  Vlasov-Uehling-Uhlenbeck theory of medium energy
      heavy ion reactions:  role of mean field dynamics and two body
collisions.
      {\em Phys.\ Rev.\ C} {\bf 31}, 1770.

C. Mader and W. Bauer (1992).  Work in progress.

Odyniec, G. \etal\ (1988).  In: Proceedings of the 8th High Energy Heavy
      Ion Study, Berkeley, edited by J. Harris and G. Wozniak,
      LBL Report 24580, p.\ 215

Ogilvie, C.A. \etal\ (1990).  The disappearance of flow and its relevance
      to nuclear matter physics.
      {\em Phys.\ Rev.\ C} {\bf 42}, R10.

Ono, A. \etal\ (1992).
      {\em Prog.\ of Theor.\ Phys.} {\bf 87}, 1185.

Peilert, G. \etal\ (1992).  Dynamical treatment of Fermi motion in a
      microscopic description of heavy ion reactions.
      {\em Phys.\ Rev.\ C} {\bf 46}, 1457.

Pluta, J. \etal\ (1992).  Work in preparation.

Pratt, S. (1984). Pion interferometry for exploding sources.
      {\em Phys.\ Rev.\ Lett.} {\bf 53}, 1219.

Pratt, S. (1986). Pion interferometry of the quark-gluon plasma.
      {\em Phys.\ Rev.\ D} {\bf 33}, 1314.

Randrup, J. and Remaud, B. (1990). Fluctuations in one-body dynamics.
      {\em Nucl.\ Phys.\ A} {\bf 514}, 339.

Randrup, J. (1992). Fluctuations and correlations in nuclear one-body
      dynamics. {\em Nucl.\ Phys.\ A} {\bf 545}, 47c.

Reinhard, P.-G. and Suraud, E. (1992). Stochastic TDHF and large fluctuations.
      {\em Nucl.\ Phys.\ A} {\bf 545}, 59c.

Schuck, P. \etal\ (1989). Semiclassical and phase space approaches to dynamical
      and collisional problems of nuclei.
      {\em Prog.\ Part.\ Nucl.\ Phys.} {\bf 22}, 181.

Shuryak, E.V. (1973). The correlations of identical pions in multibody
      production.  {\em Phys.\ Lett.} {\bf B44}, 387.

St\"ocker, H. and Greiner, W. (1986). High energy heavy ion collisions -
      probing the equation of state of highly excited hadronic matter.
      {\em Phys.\ Rep.} {\bf 137}, 277.

Stock, R. \etal\ (1982). Compression effects in relativistic nucleus-nucleus
      collisions.  {\em Phys.\ Rev.\ Lett.} {\bf 49}, 1236.

Wang, S.J. \etal (1991). Relativistic transport theory for hadronic matter.
      {\em Ann.\ Phys.\ (N.Y.)} {\bf 209}, 251.

Wong, C.Y. (1982). Dynamics of nuclear fluid. VIII. Time-dependent Hartree-Fock
      approximation from a classical point of view.
      {\em Phys.\ Rev.\ C} {\bf 25}, 1460.

Zhu, F. \etal\ (1991). Light-particle correlations and the $^3$He + Ag
      reaction at 200 MeV. {\em Phys.\ Rev.\ C} {\bf 44}, R582.

}
\end{document}